\documentclass{PoS}
\usepackage{wrapfig}
\usepackage{amsmath}
\usepackage{epsfig}

\title{Measurement of Dijet Production in Diffractive Deep-Inelastic Scattering at HERA}

\ShortTitle{Dijet Production in Diffractive Deep-Inelastic Scattering}

\author{\speaker{Stefan Schmitt}\thanks{On behalf of the H1 collaboration}\\
        DESY, Notekestr.~85, 22607 Hamburg, Germany\\
        E-mail: \email{sschmitt@mail.desy.de}}

\abstract{The production of dijets is measured in diffractive
  deep-inelastic scattering at HERA. The data were recorded with the
  H1 detector at DESY in the years 2003-2007. Diffractive events are
  selected by requiring a gap in the rapidity distribution of the
  hadronic systen, where no particles are produced. Two jets are
  selected with transverse momenta in the hadronic-centre-of-mass
  system larger than $4$ and $5.5\,\text{GeV}$,
  respectively. Cross sections are measured single- and
  double-differentially in various kinematic quantities. The data are
  found to be in good agreement with NLO QCD calculations based on
  diffractive parton densities determined from inclusive diffractive
  cross section measurements.}

\FullConference{XXIII International Workshop on Deep-Inelastic Scattering,\\
		27 April - May 1 2015\\
		Dallas, Texas}

\begin{document}
\section{Introduction}

\begin{figure}[t]
\begin{center}
\epsfig{file=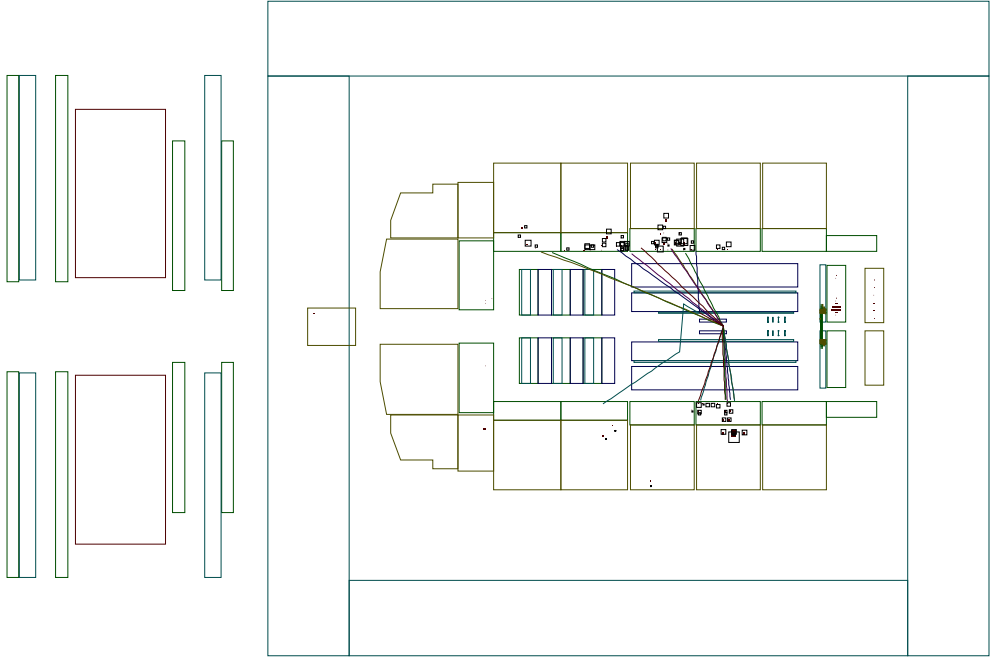,width=0.9\textwidth}
\caption{\label{fig:event}Diffractive dijet event as detected in the
  H1 experiment.}
\end{center}
\end{figure}
At HERA, reactions of electrons or positrons and protons, $ep\to eX$,
are probed at centre-of-mass energies of $320\,\text{GeV}$.
At
sufficiently large negative momentum transfer squared from the ingoing
to the outgoing electron\footnote{Throughout this paper, the term
  electron or the variable $e$ is used to denote both electrons and
  positrons, unless otherwise stated.},
$Q^2>4\,\text{GeV}$, the 
process is referred to as deep-inelastic scattering (DIS), and $Q^2$ defines a
hard scale for perturbative QCD calculations.
In the analysis presented here \cite{Andreev:2014yra},
dijet production in diffractive
DIS is studied. The reaction may be written as
$ep\to eXY$, where the system $X$ contains at least two hard jets and
the system $Y$ is a proton or a low-mass excitation, carrying a large
momentum fraction of the incoming proton. A typical event display is
shown in figure \ref{fig:event}.
The low-mass system $Y$ escapes detection to the left (forward
direction). The electron is detected in the right-most part of the
detector.
The system $X$ consists of tracks and energy
deposits in the calorimeter, forming two hard jets.
There is a large gap in
pseudorapidity\footnote{The $z$ axis is pointing along the proton
  flight direction. Polar angles $\theta$ are measured with respect to
  the $z$ axis. The pseudorapidity is defined as $\eta=-\ln\tan (\theta/2)$.} between the acceptance limit of the detector in the forward
direction and the most forward calorimeter deposit above noise level.

\begin{wrapfigure}{R}{0.3\textwidth}
\begin{center}
\epsfig{file=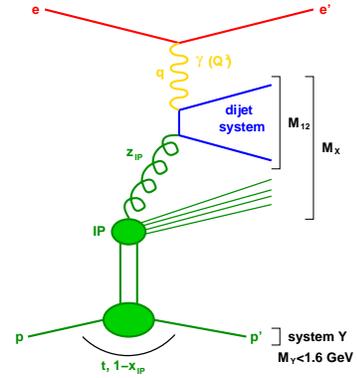,width=0.3\textwidth}
\caption{\label{fig:diffdijet}Diffractive dijet production in
  deep-inelastic scattering at HERA.}
\end{center}
\end{wrapfigure}
QCD calculations for this process are based on diffractive parton
densities (DPDFs).
The DPDFs describe the probability to find a parton with longitudinal
momentum fraction $z_{I\-\-P}$ in the proton, given that there is a
diffractive signature, characterised by a longitudinally momentum
fraction $1-x_{I\-\-P}$, momentum transfer $t$ of the outgoing proton
and a hard scale $\mu$.
They are folded with hard matrix elements to describe jet production at
next-to-leading order in the strong coupling.
Experimentally, the DPDFs are determined from inclusive diffractive
measurements, where no requirements on the hadronic system $X$
are made.
As suggested by the scheme shown in figure \ref{fig:diffdijet}, the
DPDFs are determined with the ad-hoc assumption that they factorise into a
probability to find a colourless object $I\-\-P$ in the
proton and parton density functions, ascribed to the structure of
$I\-\-P$.
The probability or ``flux factor'' is taken to depend on
$x_{I\-\-P}$ and $t$ only, 
whereas the parton density functions of $I\-\-P$ only depend on the
variable $z_{I\-\-P}$ and the hard scale $\mu$.
For the present analysis,
the H1 2006 DPDF fit B \cite{Aktas:2006hy} is used to predict cross sections.
Several measurements at HERA 
have confirmed the validity of the NLO calculations in diffractive DIS
\cite{Chekanov:2002qm,Aktas:2006up,Aktas:2007bv,Chekanov:2007aa,Aaron:2011mp,Andreev:2015cwa}.
The present analysis is exploiting the high statistics HERA II sample,
in order to obtain more precise results and enable
double-differential cross section measurements.

\section{Data analysis}

The analysis \cite{Andreev:2014yra} is based on events recorded with
the H1 detector \cite{Abt:1996hi} in the 
years 2005-2007, corresponding to an integrated luminosity of
$290\,\text{pb}^{-1}$.
DIS events are identified with an electron in
the H1 rear calorimeter (SpaCal) \cite{Appuhn:1996na}.
This limits the accessible kinematic range in
momentum transfer to $4<Q^2<100\,\text{GeV}^2$. The hadronic final
state $X$ is reconstructed from the detected tracks and calorimeter
deposits, using an energy flow algorithm. The momentum transfer
$Q^2$ and the inelasticity $y$ are reconstructed with the $e\Sigma$
method \cite{Adloff:1997sc}, which uses properties of both the
electron and the hadronic 
final state. The inelasticity is restricted to $0.1<y<0.7$.

Diffractive events are further selected with the condition of a Large
Rapidity Gap (LRG) separating  the hadronic systems $X$ and $Y$. The
LRG condition requires that the
most forward energy deposit above noise level in the H1 calorimeter has a
pseudorapidity $\eta_{\max}<3.2$ and that there are no signals in the
H1 forward muon system or in the forward tagging system. The latter
devices extend the reach for vetoing particles up to $\eta\sim
7.5$. The momentum fraction $x_{I\-\-P}=(Q^2+M_X^2)/(sy)$ is restricted to
the range $x_{I\-\-P}<0.03$. Here, $M_X$ is the invariant mass of the
hadronic system $X$, reconstructed from all detected particles after
excluding the electron. The system $Y$ is not detected; however from
energy-momentum conservation one can deduce that its mass is limited
by the diffractive selection to $M_Y\lesssim
1.6\,\text{GeV}$ and the momentum transfer at the proton vertex is
limited to $\vert t\vert\lesssim1\,\text{GeV}^2$.

The selection criteria are summarised in table \ref{tab:selection}.
\begin{table}[t]
\begin{tabular}{c|c|c}
\hline
 & Extended Analysis Phase Space & Measurement Cross Section Phase
 Space \\
\hline
DIS & $3<Q^2<100\,\text{GeV}^2$ & $4<Q^2<100\,\text{GeV}^2$ \\ 
    & $y<0.7$ & $0.1<y<0.7$ \\
\hline
Diffraction & $x_{I\-\-P}<0.04$ & $x_{I\-\-P}<0.03$ \\
& LRG requirements & $\vert t\vert<1\,\text{GeV}^2$ \\
& & $M_Y<1.6\,\text{GeV}$ \\
\hline
Dijets & $p_{T,1}^{\star}>3\,\text{GeV}$ &
$p_{T,1}^{\star}>3\,\text{GeV}$ \\
 & $p_{T,1}^{\star}>5.5\,\text{GeV}$ &
$p_{T,1}^{\star}>4\,\text{GeV}$ \\
 & $-2<\eta_{1,2}^{\text{lab}}<2$ &  $-1<\eta_{1,2}^{\text{lab}}<2$ \\
\hline
\end{tabular}
\caption{\label{tab:selection}Analysis phase space.}
\end{table}
The final state $X$ is boosted to the $\gamma^{\,\star}\-p$
restframe and jets are reconstructed using the inclusive $k_T$ jet
algorithm \cite{Catani:1992zp} with $P_T$ recombination scheme and
distance parameter $R=1$.
Dijet events are accepted if there are at least two jets with
transverse momenta of the leading (subleading) jet fulfilling the conditions
$p_{T,1}^{\star}>5.5\,\text{GeV}$ ($p_{T,2}^{\star}>4\,\text{GeV}$). The jet pseudorapidity is
restricted in the laboratory frame to $-1<\eta_{1,2}^{\text{lab}}<2$, in
order to ensure that the jets are well contained.

\begin{wrapfigure}{R}{0.53\textwidth}
\begin{center}
\epsfig{file=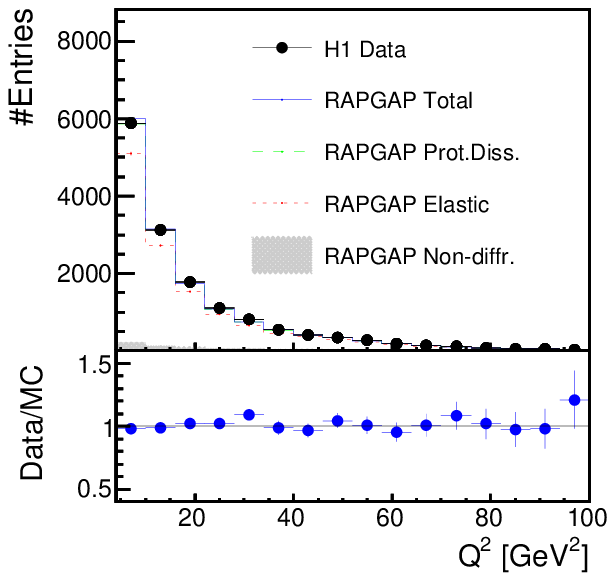,width=0.22\textwidth}%
\epsfig{file=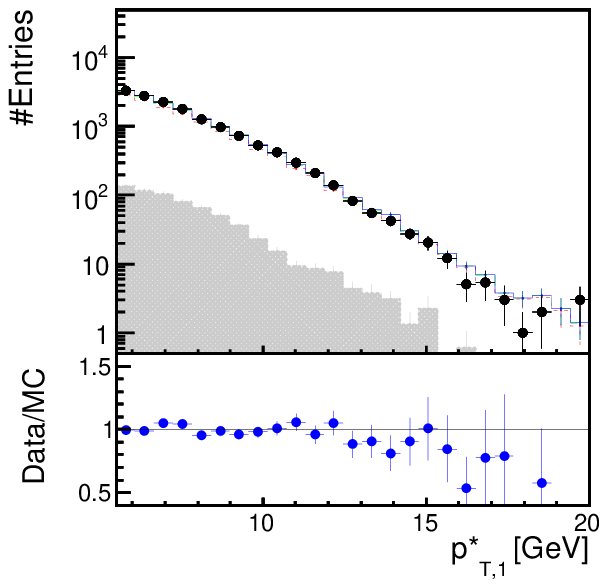,width=0.22\textwidth}
\epsfig{file=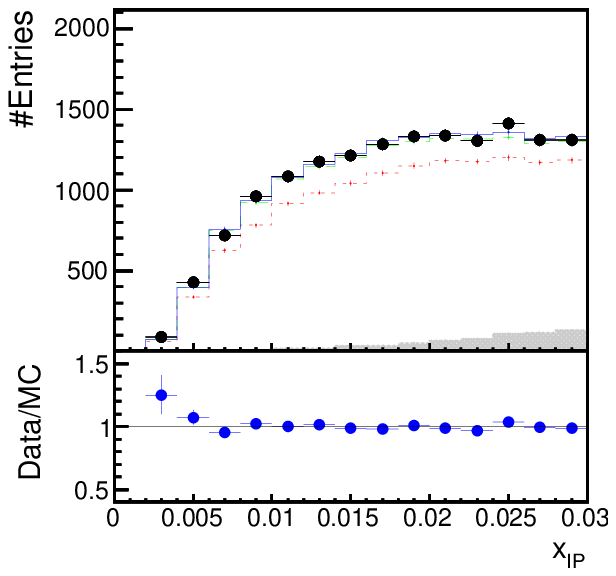,width=0.22\textwidth}%
\epsfig{file=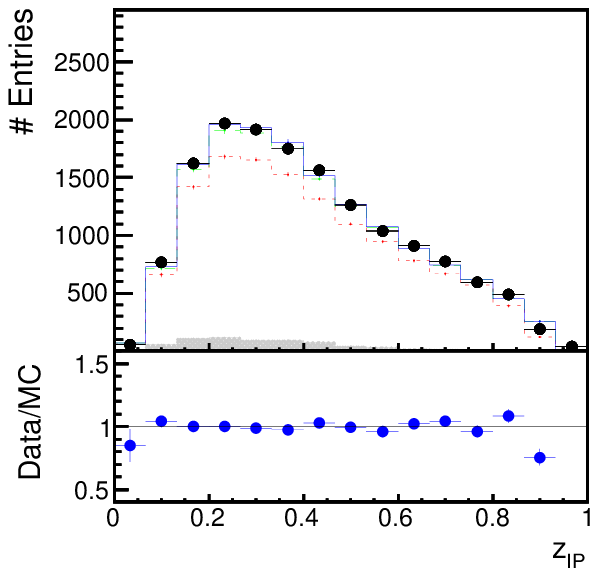,width=0.22\textwidth}
\caption{\label{fig:control}Control distributions of selected
  kinematic variables at detector level.}
\end{center}
\end{wrapfigure}
The events are selected in an extended analysis phase space, in
order to ensure that migrations near the phase space boundaries can be
well controlled.
After correcting for detector effects, the phase space
is restricted to the boundaries described in the text.
For unfolding from detector objects to the particle level, a
regularised unfolding technique is applied.
Typically, the distributions are measured in twice as many bins as are
unfolded later.
Choosing more bins reconstructed than measured later, improves the
unfolding results and reduces the correlations between data points.
The resulting correlation coefficients are typically close to
zero and do not exceed the range $\vert\rho_{ij}\vert <0.6$. In fact, only
for a few cases, where the detector resolution is limited,
correlations coefficients $\vert\rho_{ij}\vert$ close to $0.6$ are observed.
In addition to the fine resolution on detector level, there are extra
bins ascribed to control regions outside the measurement phase space.
The extra bins are also unfolded such that the normalisation of the
prediction outside measurement phase space is taken from data.
This procedure enhances the stability of the analysis to systematic
effects related to the RAPGAP prediction and its reweighting.

Control distributions of reconstructed variables at detector level are
shown in figure \ref{fig:control}. These are the momentum transfer
$Q^2$, the leading jet transverse momentum, and the diffractive
momentum fractions $x_{I\-\-P}$ and $z_{I\-\-P}$, where
$z_{I\-\-P}=(Q^2+M_{12}^2)/(Q^2+M_X^2)$ and $M_{12}$ is the invariant mass
of the dijet system. The variables are all well described by the
prediction, which is used to describe migration effects
between particle level and detector objects for the unfolding
procedure.
The prediction is evaluated using the RAPGAP event generator and a
GEANT-based simulation of the H1 detector.
It has been reweighted in three variables to improve the description;
figure \ref{fig:control} shows the variables after reweighting.
\section{Single-differential cross sections}

Single-differential cross sections for
diffractive dijet production in DIS are measured
in the phase-space defined in table \ref{tab:selection}. The dominant
systematic uncertainties are: the hadronic energy scale uncertainty of
$1\%$, model uncertainties of the RAPGAP prediction and normalisation
uncertainties related to the LRG selection.
The normalisation uncertainties amount to about $7\%$, the model
uncertainties have a size of typically $5\%$ and the hadronic energy
scale causes uncertainties on the cross section of typically $4\%$.

\begin{figure}[t]
\begin{center}
\epsfig{file=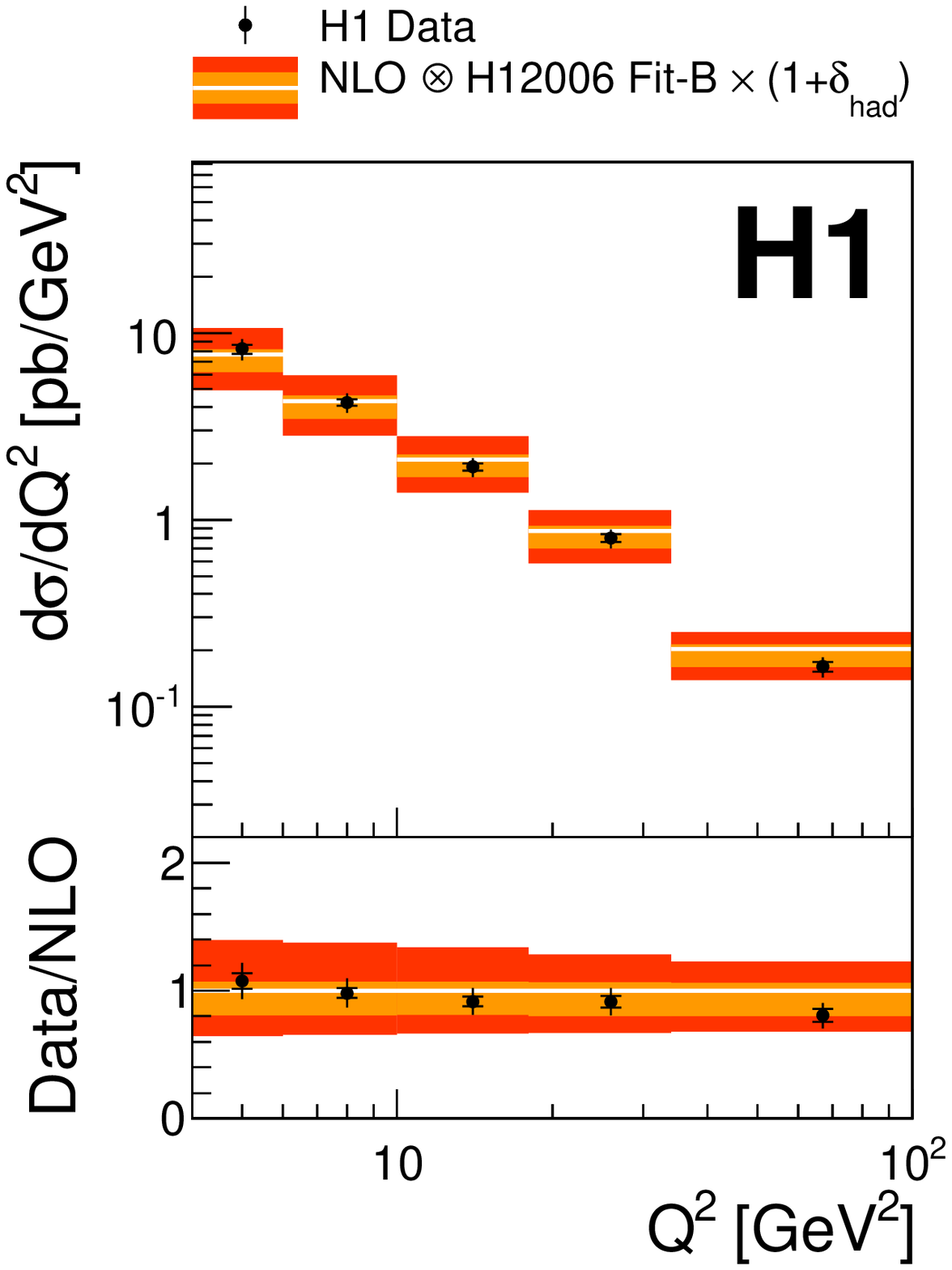,width=0.24\textwidth}%
\epsfig{file=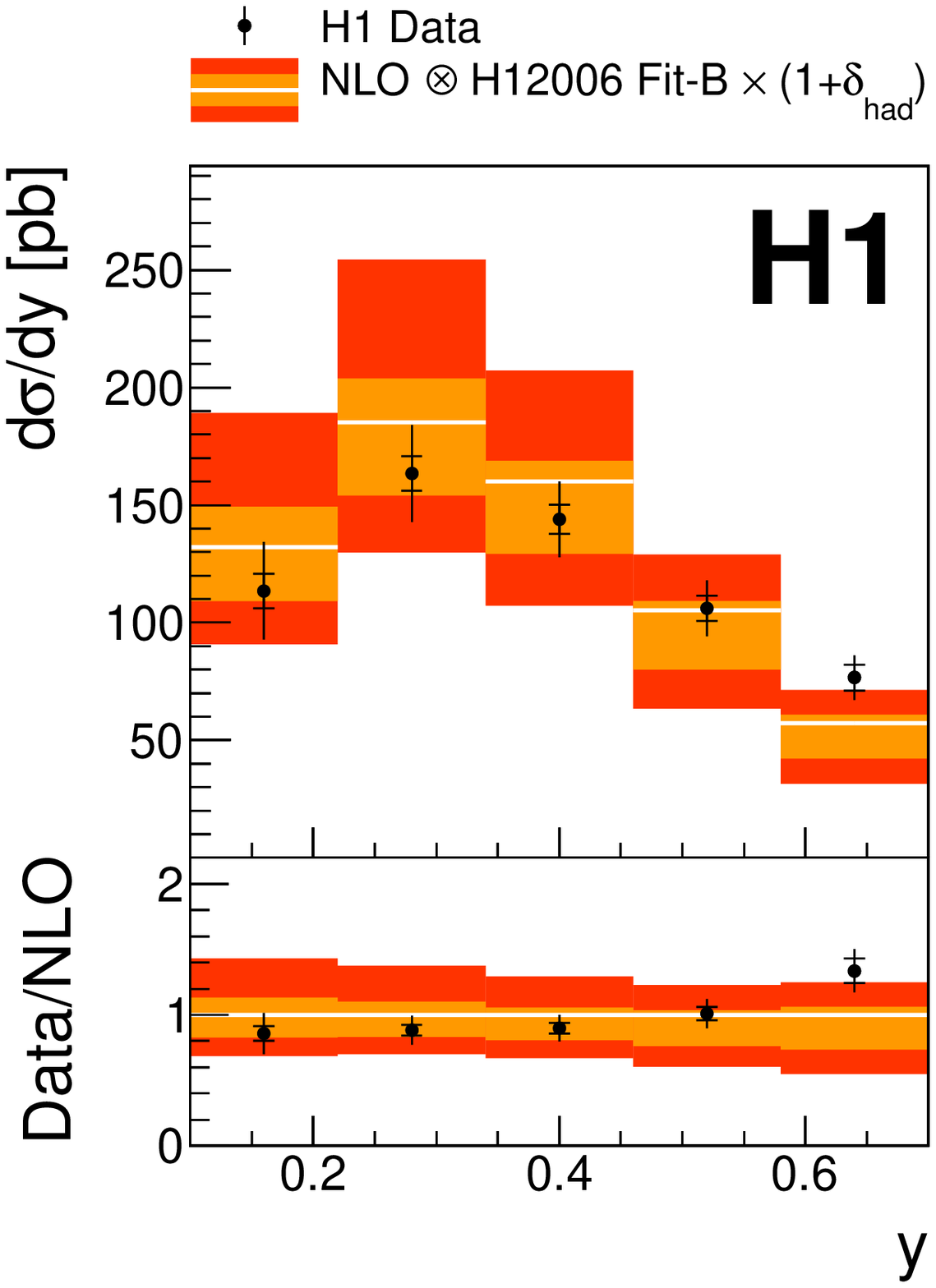,width=0.24\textwidth}%
\epsfig{file=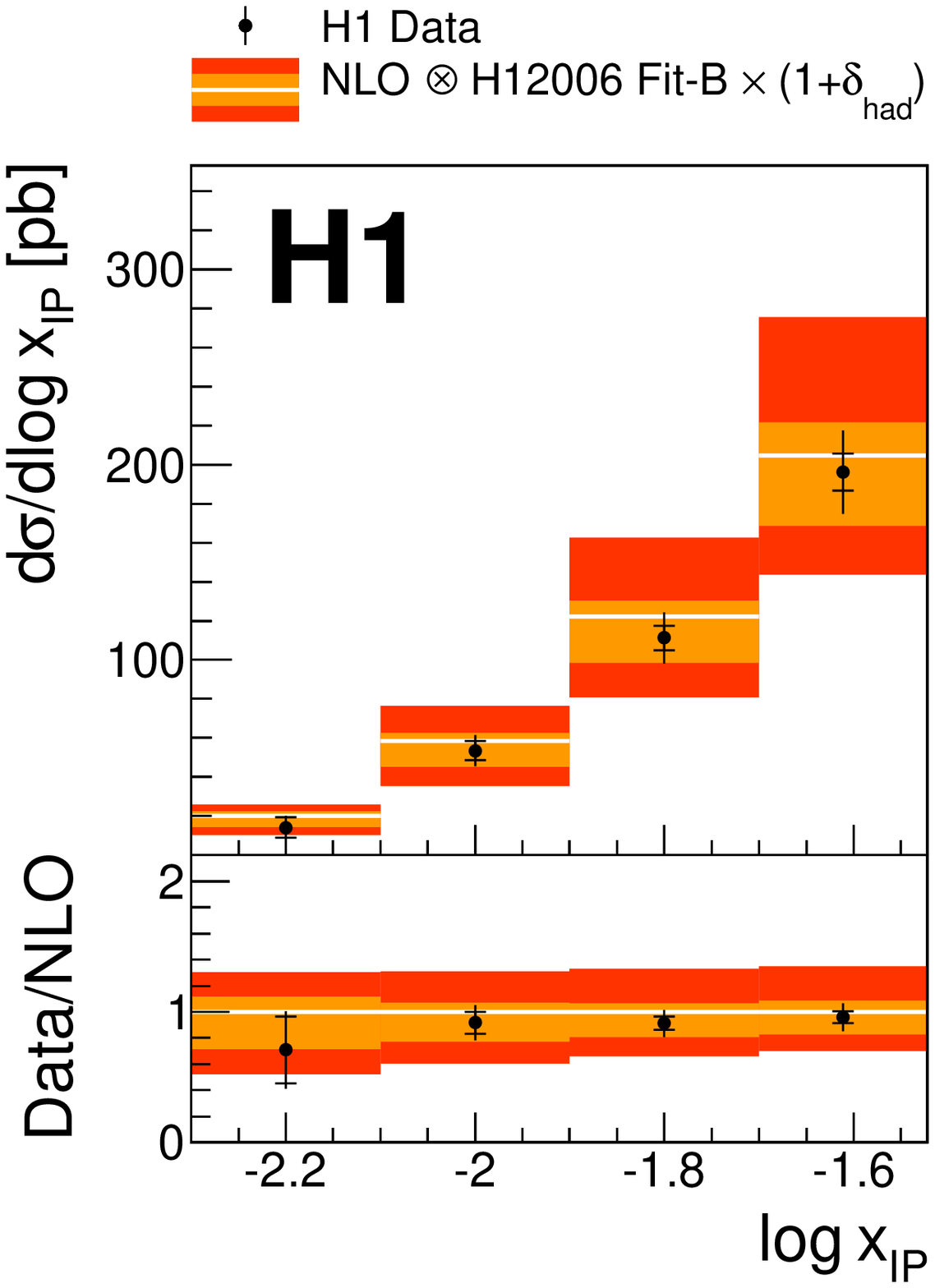,width=0.24\textwidth}%
\epsfig{file=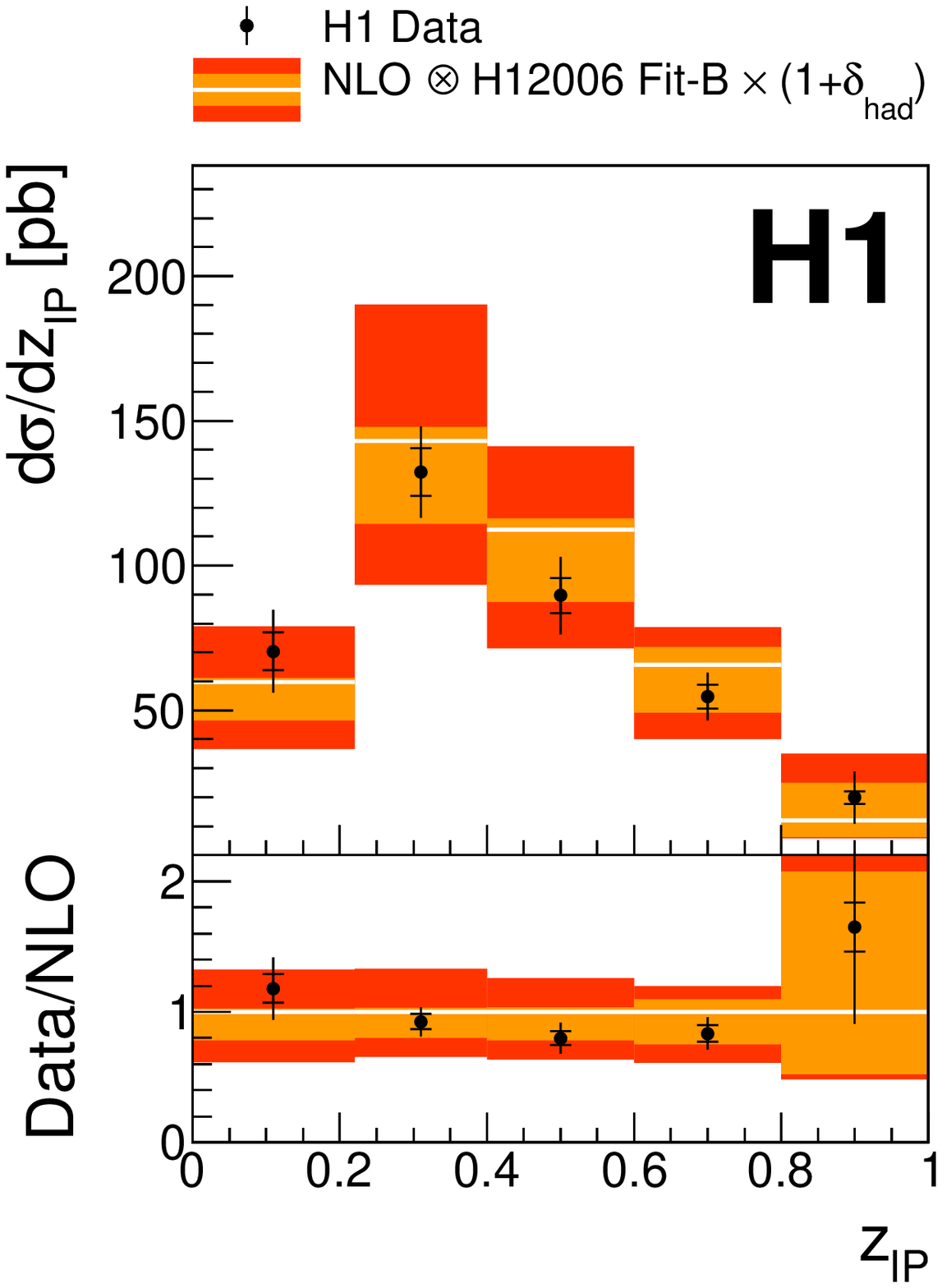,width=0.24\textwidth}
\caption{\label{fig:single1}Single-differential cross sections for
  diffractive dijet production in DIS as a function of the variables
  $Q^2$, $y$, $x_{I\-\-P}$ and $z_{I\-\-P}$.}
\end{center}
\end{figure}
\begin{figure}[t]
\begin{center}
\epsfig{file=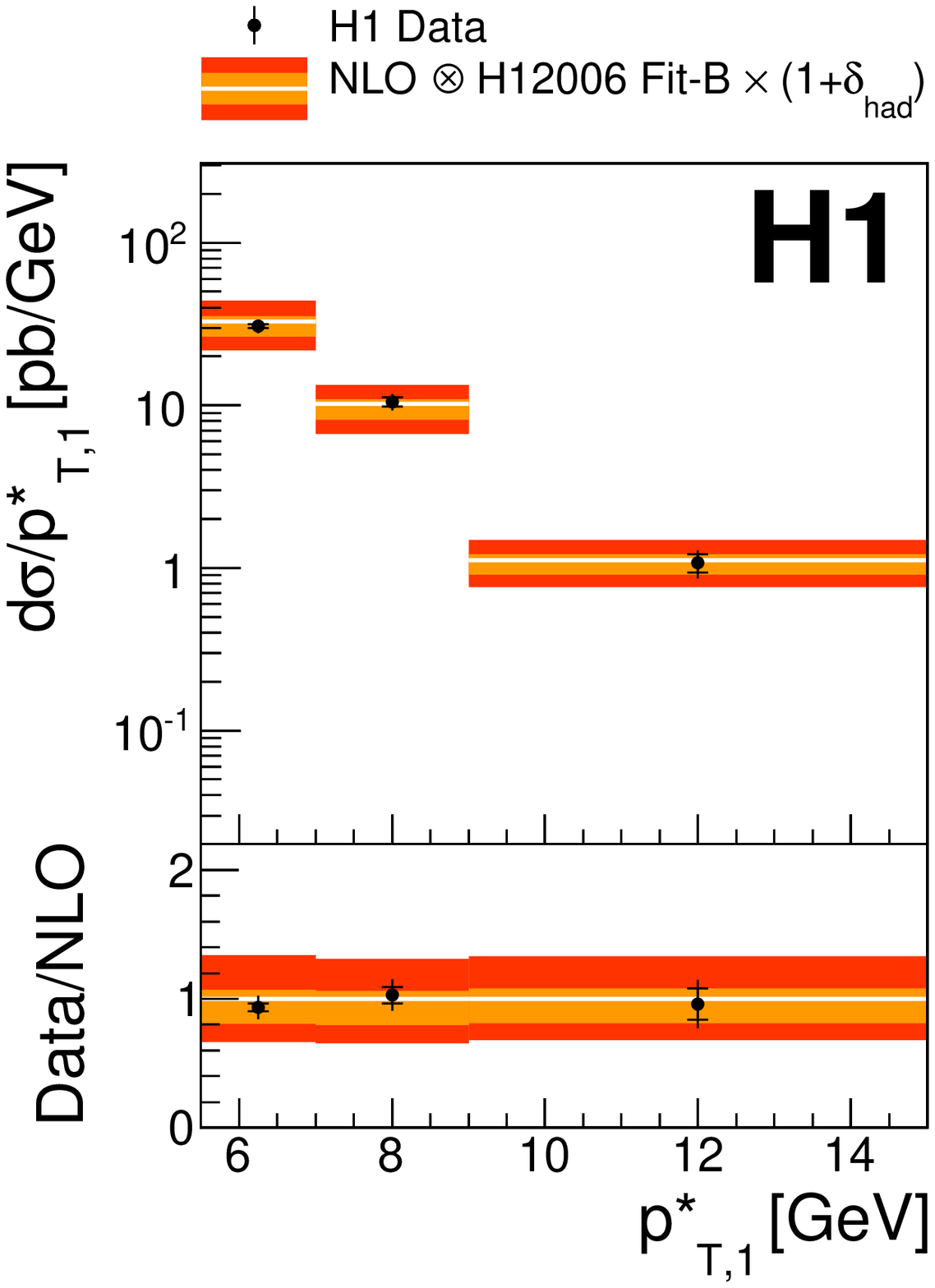,width=0.24\textwidth}%
\epsfig{file=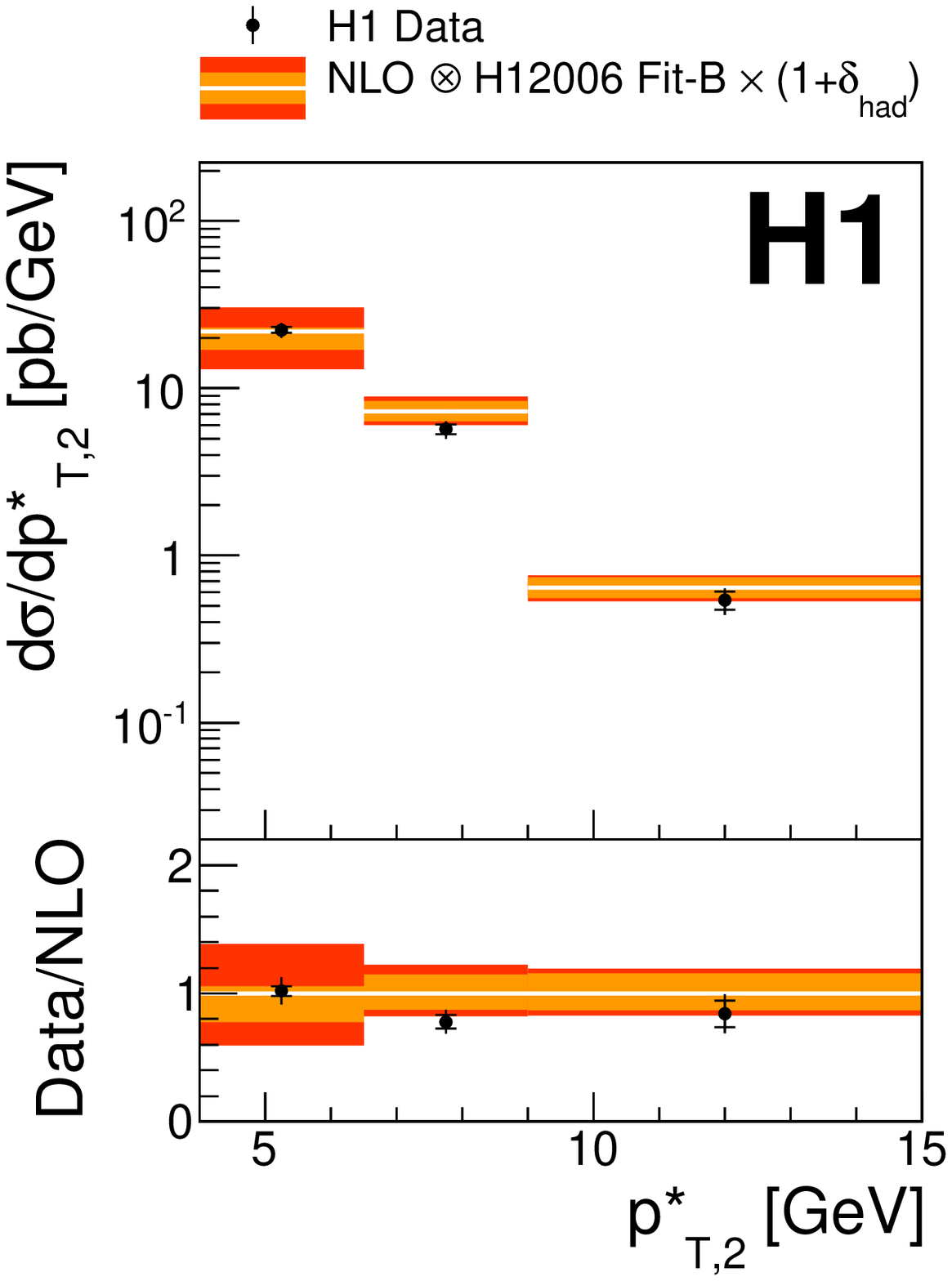,width=0.24\textwidth}%
\epsfig{file=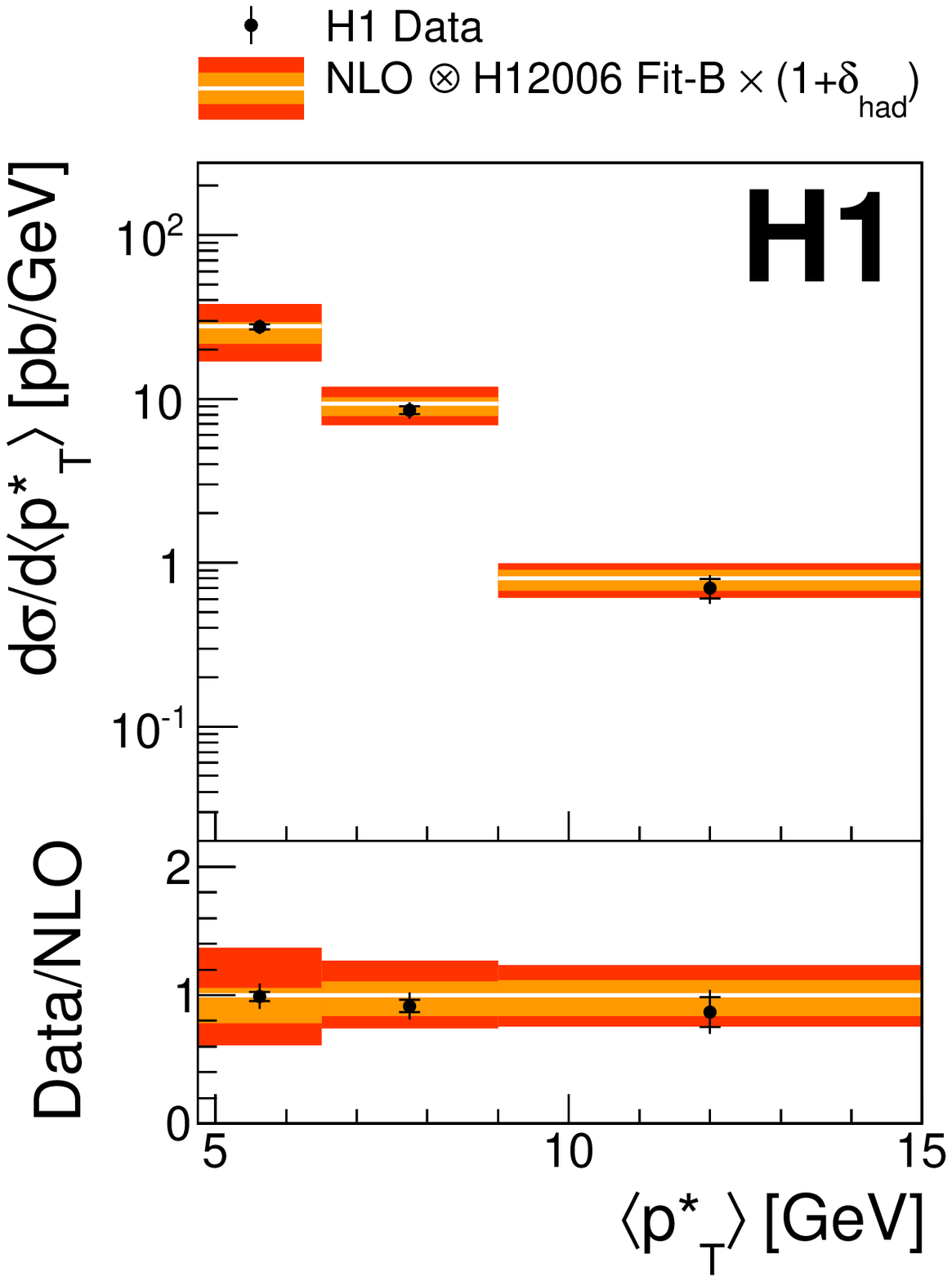,width=0.24\textwidth}%
\epsfig{file=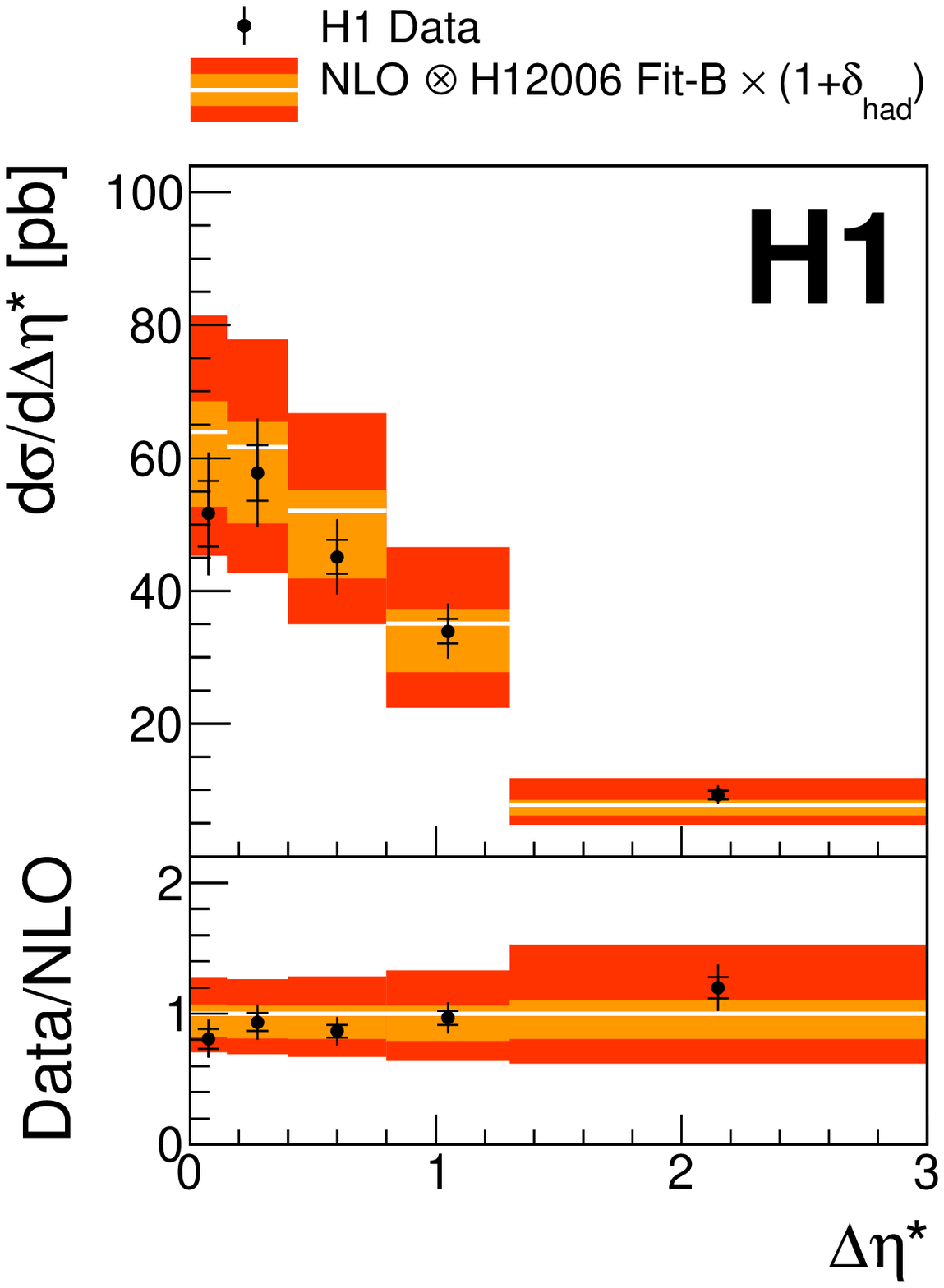,width=0.24\textwidth}
\caption{\label{fig:single2}Single-differential cross sections for
  diffractive dijet production in DIS as a function of the variables
  $p_{T,1}^{\star}$, $p_{T,2}^{\star}$, $\langle p_T\rangle$ and $\Delta\eta^{\star}$.}
\end{center}
\end{figure}
Single differential cross sections are measured as a function of
the variables $Q^2$, $y$, $\log x_{I\-\-P}$, $z_{I\-\-P}$, $p_{T,1}$,
$p_{T,2}$, the average transverse jet momentum $\langle
p_T\rangle=(p_{T,1}^{\star}+p_{T,2}^{\star})/2$
and the difference in pseudorapidity
$\Delta\eta^{\star}=\vert\eta_1-\eta_2\vert$. The results are shown
in figure \ref{fig:single1} and \ref{fig:single2}.
The data points are shown with their statistical and total uncertainties.
The precision of the cross section measurements is limited by
systematic effects in most cases.
They are compared to NLO QCD predictions, obtained using the NLOJET++
program \cite{Nagy:1998bb,Nagy:2001xb}
The predictions are shown
with their DPDF uncertainties and their total uncertainties, then also
including scale uncertainties.
In general, the data precision is better than the precision of the prediction.
In most cases the data precision is such that it is superior to the
DPDF uncertainties, so the data have the potential to improve the DPDF
fits.
This is most evident for the $z_{I\-\-P}$ distribution. It is worth to
note that no data with $z_{I\-\-P}>0.8$ has been included in the
determination of the DPDF fit at the time.
However, because the NLO predictions suffer from sizable scale
uncertainties, it is difficult to draw further conclusions or to
pursue a DPDF fit, unless NNLO calculations for jet production in
DIS become available.

\section{Double-differential cross sections and extraction of the
  strong coupling}

\begin{figure}[t]
\begin{center}
\epsfig{file=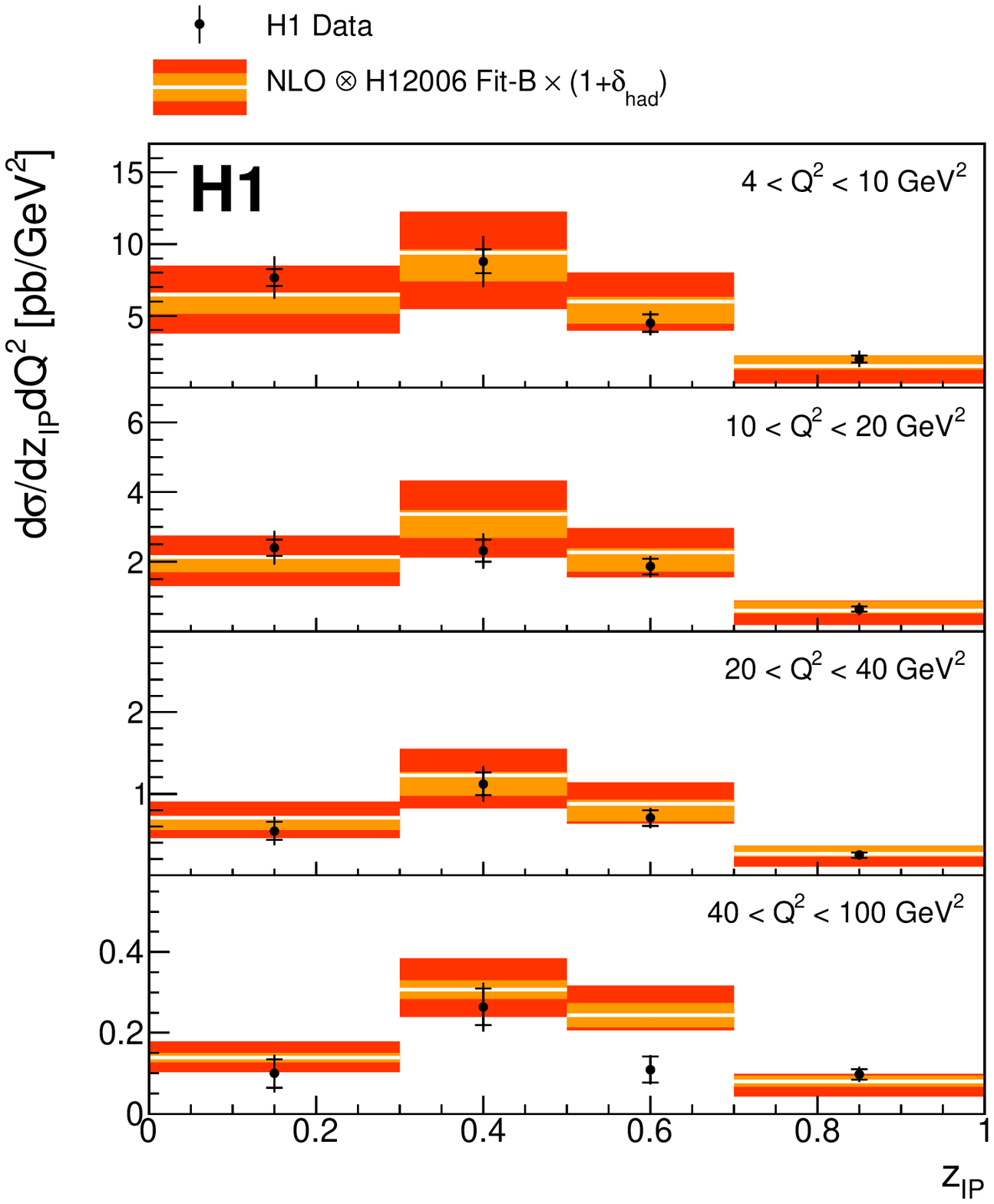,width=0.47\textwidth}%
\epsfig{file=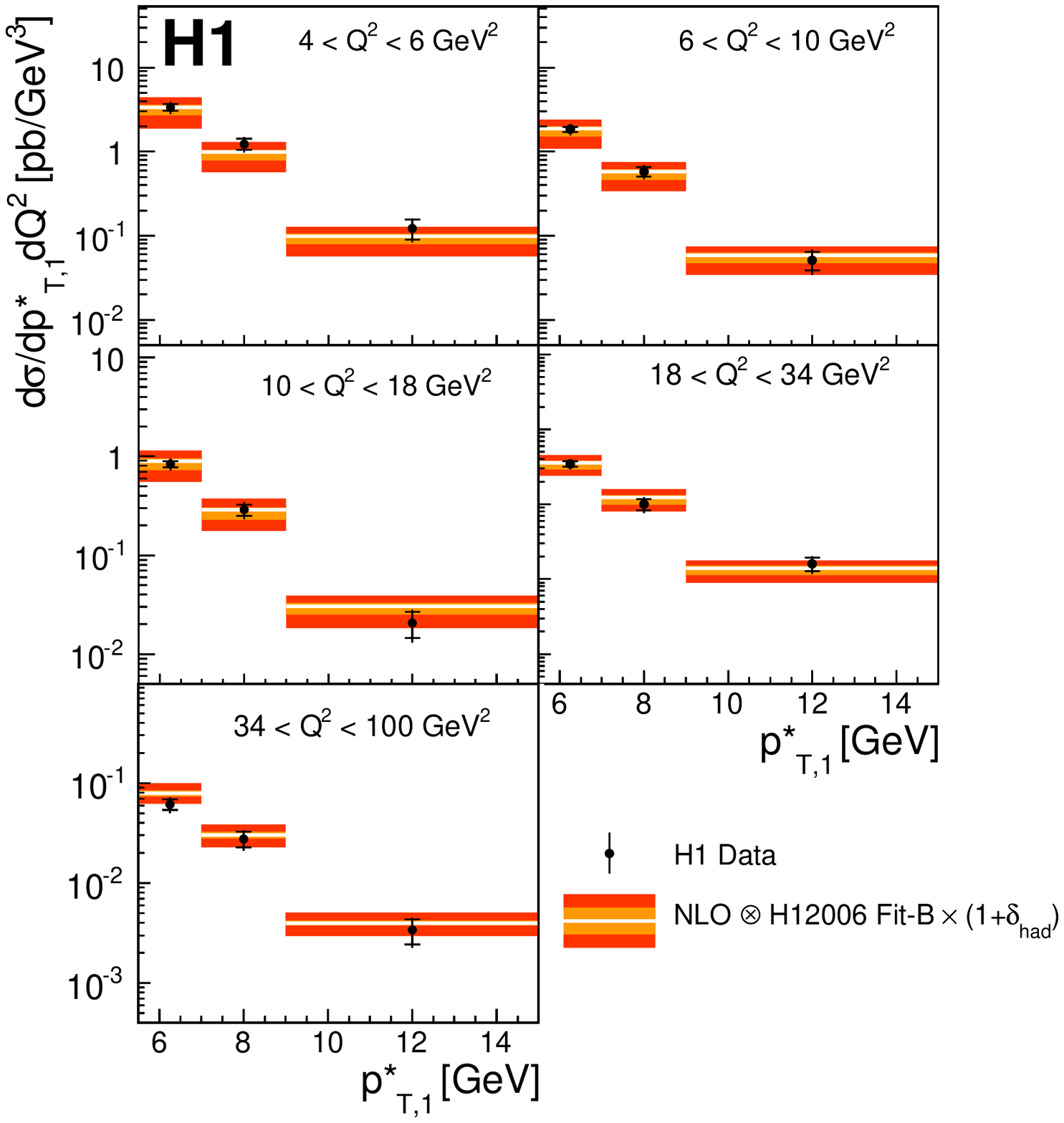,width=0.49\textwidth}
\caption{\label{fig:double}Double-differential cross sections for
  diffractive dijet production in DIS as a function of the variables
  $Q^2$ and $z_{I\-\-P}$ and as a function of the variables $Q^2$
  and $p_{T,1}^{\star}$.}
\end{center}
\end{figure}
Double-differential cross sections are measured as a function of the
variables $Q^2$ and $z_{I\-\-P}$ and as a function of the variables $Q^2$
and $p_{T,1}^{\star}$. The results are shown in figure
\ref{fig:double}.
Similar to the case of single-differential cross sections, the data
precision is not limited by the statistical errors in most cases.
The data are well described by the NLO QCD predictions, with the
exception of one point in the ($Q^2,z_{I\-\-P}$) distribution. At high
$z_{I\-\-P}>0.5$ the data are more precise than the prediction for any
choice of $Q^2$.
In order to evaluate the sensitivity of the data to parameters of the
QCD prediction, a fit of the strong coupling constant $\alpha_s$ is
performed, using the cross sections measured double-differentially
in $Q^2$ and $p_{T,1}^{\star}$.
The fit yields $\alpha_s(M_Z) = 0.119 \pm 0.004 (exp) \pm 0.002 (had)
\pm 0.005 (DPDF) \pm 0.010 (\mu_r) \pm 0.004 (\mu_f ) (11)$.
%
This first determination of $\alpha_s$ in hard diffraction at HERA is
carried out to as a consistency check and to evaluate the level of
experimental as compared to theoretical uncertainties.
The experimental precision of $\pm 0.004$ is better than the
uncertainties stemming from the knowledge of DPDFs $\pm0.005$.
The largest uncertainties are related to limitations of the NLO
calculation, $\pm 0.010$ and $\pm 0.004$ from the variation of the
renormalisation and factorisation scales, respectively.

\section{Summary}

Diffractive dijet production in DIS is measured
at HERA, using data recorded with the H1 detector. As compared to
previous measurements on smaller data samples, the precision has been
improved significantly, and is now limited by systematic
effects. Single- and double-differential cross sections are determined
for a selection of variables.
The data are consistent with NLO QCD calculations based on
DPDF fits from inclusive diffractive data. This confirms that
factorisation holds in diffractive deep-inelastic scattering. Using
the NLO calculations, the strong coupling is extracted for the first
time in hard diffraction at HERA, and found to be consistent with
other $\alpha_s$ measurements. The NLO QCD predictions, however,
suffer from large scale uncertainties. Further interpretation of the
data will benefit from NNLO calculations of jet production in
deep-inelastic scattering.


\begin{thebibliography}{99}
\bibitem{Andreev:2014yra}
  V.~Andreev {\it et al.} [H1 Collaboration],
  JHEP {\bf 1503} (2015) 092
  [arXiv:1412.0928 [hep-ex]].

\bibitem{Aktas:2006hy}
  A.~Aktas {\it et al.} [H1 Collaboration],
  Eur.\ Phys.\ J.\ C {\bf 48} (2006) 715
  [hep-ex/0606004].

\bibitem{Chekanov:2002qm}
S.~Chekanov {\em et~al.}  [ZEUS Collaboration],
  \href{http://dx.doi.org/10.1016/S0370-2693(02)02595-9}{{\em Phys.Lett.}
  {\bfseries B545} (2002) 244--260},
\href{http://arxiv.org/abs/0206020}{{\ttfamily [hep-ex/0206020]}}.

\bibitem{Aktas:2006up}
A.~Aktas {\em et~al.}  [H1 Collaboration],
  \href{http://dx.doi.org/10.1140/epjc/s10052-006-0206-2}{{\em Eur.Phys.J.}
  {\bfseries C50} (2007) 1--20},
\href{http://arxiv.org/abs/0610076}{{\ttfamily [hep-ex/0610076]}}.

\bibitem{Aktas:2007bv}
A.~Aktas {\em et~al.}  [H1 Collaboration],
  \href{http://dx.doi.org/10.1088/1126-6708/2007/10/042}{{\em JHEP} {\bfseries
  10} (2007) 042},
\href{http://arxiv.org/abs/0708.3217}{{\ttfamily [arXiv:0708.3217]}}.

\bibitem{Chekanov:2007aa}
S.~Chekanov {\em et~al.}  [ZEUS Collaboration],
  \href{http://dx.doi.org/10.1140/epjc/s10052-007-0426-0,
  10.3360/dis.2007.112}{{\em Eur.Phys.J.} {\bfseries C52} (2007) 813--832},
\href{http://arxiv.org/abs/0708.1415}{{\ttfamily [arXiv:0708.1415]}}.

\bibitem{Aaron:2011mp}
F.~Aaron {\em et~al.}  [H1 Collaboration],
  \href{http://dx.doi.org/10.1140/epjc/s10052-012-1970-9}{{\em Eur.Phys.J.}
  {\bfseries C72} (2012) 1970},
\href{http://arxiv.org/abs/1111.0584}{{\ttfamily [arXiv:1111.0584]}}.

\bibitem{Andreev:2015cwa}
  V.~Andreev {\it et al.} [H1 Collaboration],
  JHEP {\bf 1505} (2015) 056
  [arXiv:1502.01683 [hep-ex]].

\bibitem{Abt:1996hi}
  I.~Abt {\it et al.} [H1 Collaboration],
  Nucl.\ Instrum.\ Meth.\ A {\bf 386} (1997) 310.

\bibitem{Appuhn:1996na}
  R.~D.~Appuhn {\it et al.} [H1 SPACAL Group Collaboration],
  Nucl.\ Instrum.\ Meth.\ A {\bf 386} (1997) 397.

\bibitem{Adloff:1997sc}
  C.~Adloff {\it et al.} [H1 Collaboration],
  Z.\ Phys.\ C {\bf 76} (1997) 613
  [hep-ex/9708016].

\bibitem{Catani:1992zp}
  S.~Catani, Y.~L.~Dokshitzer and B.~R.~Webber,
  Phys.\ Lett.\ B {\bf 285} (1992) 291.

\bibitem{Nagy:1998bb}
  Z.~Nagy and Z.~Trocsanyi,
  Phys.\ Rev.\ D {\bf 59} (1999) 014020
  [hep-ph/9806317],
   Erratum-ibid.~{\bf 62} (2000) 099902.

\bibitem{Nagy:2001xb}
  Z.~Nagy and Z.~Trocsanyi,
  Phys.\ Rev.\ Lett.\  {\bf 87} (2001) 082001
  [hep-ph/0104315].
  
\end{thebibliography}
\end{document}